\documentclass[preprint2]{aastex}
\shorttitle{8GHz Survey}
\shortauthors{Kellermann et al.}
\begin{document}

\title{The Micro-Jansky Sky at 8.4 GHz}

\author{E.~B.~Fomalont and K.\ I.\ Kellermann}
\affil{National Radio Astronomy Observatory, Charlottesville, VA 22903}
\email{efomalon@nrao.edu, kkellerm@nrao.edu}

\author{R.~ B.~Partridge}
\affil{Haverford College, Haverford, PA 19041}
\email{bpartrid@haverford.edu}

\author{R. A. Windhorst}
\affil{Arizona State University, Tempe, AZ 85287}
\email{Rogier.Windhorst@asu.edu}

\author{E.~A.~Richards}
\affil{Department of Astronomy, University of Alabama, Huntsville, AL 35899}
\email{eric.richards@msfc.nasa.gov}

\begin{abstract}
We present the results from two radio integrations at 8.4 GHz using
the VLA.  One of the fields, at 13$^h$+43$^\circ$ (SA13 field), has an
rms noise level of $1.49~\mu$Jy and is the deepest radio image yet
made.  Thirty-four sources in a complete sample were detected above
$7.5~\mu$Jy and 25 are optically identified to a limit of I=25.8,
using our deep HST and ground-based images.  The radio sources are
usually located within $0.5''$ (typically 5 kpc) of a galaxy nucleus,
and generally have a diameter less than $2.5''$.  The second field at
17$^h$+50$^\circ$ (Hercules Field) has an rms noise of $35~\mu$jy and
contains 10 sources.

We have also analyzed a complete flux density-limited sample at 8.4
GHz of 89 sources from five deep radio surveys, including the Hubble
deep and flanking fields as well as the two new fields.  Half of all
the optical counterparts are with galaxies brighter than I=23 mag, but
20\% are fainter than I=25.5 mag.  We confirm the tendency for the
micro-Jansky radio sources to prefer multi-galaxy systems.

The distribution of the radio spectral index between 1.4 and 8.4 GHz
peaks at $\alpha\approx -0.75~(S\sim\nu^{+\alpha}$), with a median
value of $-0.6$.  The average spectral index becomes steeper (lower
values) for sources below $35~\mu$Jy, and for sources identified with
optical counterparts fainter than I=25.5 mag.  This correlation
may suggest that there is an increasing contribution from star-burst
galaxies compared to active galactic nuclei (AGNs) at lower radio
flux densities and fainter optical counterparts.

The differential radio count between 7.5 and 1000 $\mu$Jy has a slope
of $-2.11\pm 0.13$ and a surface density of 0.64 sources
(arcmin)$^{-2}$ with flux density greater than $7.5~\mu$Jy.
\end{abstract}
\section{Introduction}
 
Deep surveys of the extragalactic sky have been made at X-ray
(Hasinger et al.\ 1998; Cowie et al.\ 2001), optical (Lilly et al.\
1996; Madau et al.\ 1996; Steidel et al.\ 1999), infra-red (Goldschmidt
et al.\ 1997; Rowan-Robinson et al.\ 1997; Elbaz et al.\ 1999), sub-mm
(Hughes et al.\ 1998; Barger et al.\ 1999; Blain et al.\ 1999; Barger
et al.\ 2000; Scott et al.\ 2000), as well as radio wavelengths
(Windhorst et al.\ 1984; Windhorst et al.\ 1985; Condon \& Mitchell
1984; Donnelly et al.\ 1987; Fomalont et al.\ 1991; Windhorst et al.\
1993; Richards et al.\ 1998; Richards et al.\ 1999; Prandoni et al.\
2001).  Much of the interest in these surveys of weak objects concern
the evidence for star formation and AGN evolution in galaxies in the
early universe, at redshifts in the range 1 to 3 (Condon \& Yin 1990;
Rowan-Robinson et al.\ 1993; Cram 1998; Mobasher et al.\ 1999; Steidel
et al.\ 1999; Haarsma et al.\ 2000).

     The radio observations are unique in that at radio wavelengths it
is possible to peer through the gas and dust that obscures the nuclear
regions of galaxies at other wavelengths.  The radio observations also
have sufficient angular resolution to distinguish between emission
that is driven by star formation and that driven by AGN.  Although
star forming activity in the early universe is perhaps most readily
observed at sub-millimeter wavelengths, current sub-millimeter
instruments do not have sufficient angular resolution to avoid
confusion due to blending of nearby sources, so high resolution radio
observations are needed to uniquely identify optical counterparts.
Moreover, because of the large negative k-correction, dusty galaxies
observed at sub-mm wavelengths all have high redshifts, whereas the
radio observations are sensitive to a wide range of redshift and a mix
of star-burst and AGN activity.

As the most sensitive radio telescope available for these kinds of
observations, the Very Large Array (VLA) has been used for many deep
surveys \citep{con84,win84,win85}.  Deep radio source surveys have
also been made with the Australian Telescope Compact Array
\citep{hop99,wil00,pra01}, and with the Westerbork Synthesis Radio
Telescope \citep{gar00}, but with less sensitivity and poorer
resolution than possible with the VLA.  These observations show that
below levels of about 10 mJy, the number of radio sources increases
more rapidly than the number between 1 mJy and 500 mJy and are
composed of a different population of sources.

 In this paper we report on the results of new 8.4 GHz VLA surveys
which cover two fields, each of diameter $9.2'$ to the 8\% sensitivity
level of the on-axis position.  One field, located at
$13^h$+43$^\circ$, which we will designate as the SA13 field, was one
of the Hubble Space Telescope Medium Deep Survey (MDS) key projects,
observed in 1992-3 \citep{gri94, win94, win95}.  These optical
observations, which were made before HST refurbishment, reach a
limiting sensitivity of I=25.5 mag.  The extensive complementary VLA
observation detected sources as faint as $7.5~\mu$Jy at 8.4 GHz.  A
preliminary account of the HST identifications in this field
\citep{win95}, hereinafter designated as Paper I, was based on the
first half of these observations which were made at lower resolution,
and included only radio sources located in the small 2.5 (arcmin)$^2$
field of view of the HST.  We have also used these same VLA
observations of this field to examine the small scale fluctuations in
the cosmic microwave background radiation \citep{par97}.  This field
has also been observed with the VLA at 1.4 GHz, covering a much larger
area, with additional KPNO optical imaging \citep{ric00, ric02}, and
with the MERLIN/VLBI National Facility of the University of Manchester
at 1.4 GHz \citep{fom02}
 
A second field, located at $17^h$+50$^\circ$ and designated at the
Hercules field, was previously imaged with WFPC2 in a 48-orbit
exposure, and reaches a 10-$\sigma$ point source sensitivity of about
27.7 mag in B and V and 26.8 in I \citep{win98}. Twelve known objects
in this field, including three radio-weak AGN, have spectroscopic
redshifts z near 2.4 (Pascarelle et al.\ 1996a; Pascarelle et al.\ 1996b;
Pascarelle et al.\ 1998; Pascarelle et al.\ 2001).  A WFPC2
image made with a medium band redshifted Ly$\alpha$ filter indicates
the presence of a number of other compact galaxy candidates with
redshifts near 2.4.  The VLA image of this field reached a
limiting flux density of only $35~\mu$Jy, but provides improved
statistics for radio sources above this level.

In $\S$2 of this paper we describe the VLA observations, calibrations
and imaging for the two fields.  The radio and optical parameters for
the sources are given in $\S$3, and detailed radio/optical images with
discussions are given in $\S$4, along with discussions of the
identifications.  In addition, using data from five deep 8.4-GHz
surveys, we discuss the spectral properties, the source density and
the optical properties of the micro-Jansky radio sources.  Further
discussion is given in the final section.

\section{The VLA Observations, Calibration and Imaging}

    The SA13 field was selected from among the HST Medium Deep Survey
fields to be devoid of strong radio sources, which would limit the
sensitivity of a deep radio survey.  This field, centered at
$\alpha=13^h12^m17.4^s$ and $\delta=42^\circ 38'05''$, epoch J2000,
was observed with the VLA for a total of 190 hours, of which we
obtained 159 hours of good data (more detail given below).  These data
included 84 hours in the most compact D-configuration (maximum
baseline 1 km) observed between 1993 October and 1994 January, and 75
hours in the C-configuration (maximum baseline 3 km) observed between
1994 November and 1995 January.  The earlier discussion in Paper I of
the optical counterparts in this field was based on the 1993/1994
D-configuration observations, but with the preliminary new
C-configuration data.  The Hercules field was observed for 12 hours in
the VLA C-configuration in 1996 February.  The center of this field is
at $\alpha=17^h14^m14.78^s$ and $\delta=50^\circ 15'29.86''$, epoch
J2000.

    The observations were made at two frequencies, 8.415 GHz and 8.465
GHz, each with dual circular polarization with a bandwidth of 50 MHz.
At this frequency the full-width half-power (FWHP) beamwidth of the
VLA 25-m antennas is $5.23'$.  We examined the area out to the 8\%
power level, at a radius of $4.55'$ from the field center.
This corresponds to a solid angle of $6.5\times 10^{-6}$
sr for each field.

   Each observing session lasted 8-10 hours and was split into
30-minute segments which included 27 minutes on the survey field and 3
minutes on a nearby calibrator source.  For the SA13 field, we used the
unresolved calibrator source, J1244+4806 ($\alpha=12^h44^m49.1955^s$,
$\delta=40^\circ 48' 06.219''$; J2000), and for the Hercules field,
the calibrator source J1658+4749 ($\alpha=16^h58^m02.7789^s$,
$\delta=47^\circ 47' 49.200$; J2000).  The position grids for the
surveys are tied to these calibrators, both of which have a position
known to better than $0.1''$.  The amplitude calibration was derived
from observations of 3C286 made once per day; we assumed a flux
density of 5.19 Jy at 8.44 GHz.

     The data from each observing session were edited for occasional
short periods of interference, excessive noise in individual
correlators or other technical problems, antenna shadowing, and
inclement weather conditions.  In the SA13 field, less than one
percent of the data was lost to interference and technical problems,
but we had to discard about 30 hours of data due to poor weather,
primarily snow, during 4 of the 27 observing sessions, each of 6 to 8
hours duration.

     Both fields were imaged and deconvolved with the AIPS tasks IMAGR
using a pixel separation $0.75''$ over a field of view of $1500''$ in
order to detect radio sources well outside of the primary beam field
of view which could seriously affect the image quality
\citep{fom91,win93}.  The visibility data were weighted (AIPS
terminology of robust=1) to obtain high signal to noise with low
sidelobe levels and good resolution.  For the SA13 field which
contained both C and D configuration data, the full-width at
half-power (FWHP) resolution was $6''$.  We also made an image with a
resolution of $4''$ to help resolve possible source blends.  For the
Hercules field with only C-configuration data, the resolution was
$3.5''$.  The measured rms noise in the SA13 field was $1.49~\mu$Jy,
which makes this the most sensitive radio image yet obtained.  For the
Hercules field the rms noise was $7.52~\mu$Jy.  The strongest source
in the SA13 field has a measured map flux density of $120~\mu$Jy.
This is less than one hundred times the thermal receiver noise limit,
so dynamic range limitations are not an important factor.  The
Hercules field contained a known `bright' radio source of 6 mJy at the
field center, 53W002 \citep{win84}, which was the target of the deep
HST study \citep{win98}.  Artifacts from the 6-mJy source limited the
dynamic range of the image, but self-calibration with a solution
interval of 3 min had sufficient signal to noise to improve the
calibration, and produce an image was signal-to-noise limited

The completeness limit of a survey is that level above which there is
near certainty that any source is real.  As discussed in previous VLA
papers on radio surveys with comparable resolution, crowdedness and
field of view, a source with a peak flux density on the image
$>5.0$-$\sigma$ has a probability of $>98\%$ of being a true detection
\citep{ric98}.  This percentage includes the effects of receiver noise
and side-lobe contamination from faint sources.  The most negative
peak flux density on the image is also a good indication of the
detection limit \citep{par97}.  We found that detection levels of
$7.5~\mu$Jy and $35.0~\mu$Jy level for the SA13 and Hercules fields,
respectively, met both of the above criteria.  We also confirmed 19 of
the 20 sources detected in Paper I using similar criteria.  In order
to search for possible extended radio sources with low surface
brightness, we convolved the images to $10''$ resolution for sources
above the detection level of the highest resolution image, but above
that for a lower resolution image.  One such extended source was found
in the SA13 field (source 27 in Table 1).

\section {The Source Lists}
     
   The radio images for the entire SA13 and the Hercules fields are
shown in Figures 1 and 2, respectively.  In Figure 3 we show the image
of the central part of the SA13 field where the radio sensitivity is
greatest.  The figure shows the excellent quality of the radio image,
the precise alignment of most of the radio/optical identifications and
the obvious faintness of the identifications for some of the sources.
The sources labeled with a number are given in Table 1.  Seven Weaker
radio sources which are coincident with an optical object are labeled
with a letter and given at the end of Table 1.

   Parameters for the 50 sources labeled in the SA13 field and the 10
sources labeled in the Hercules field have entries listed in Tables 1
and 2, respectively.  Thirty-four sources in the SA13 field were found
in the complete sample with a peak flux density greater than
$7.5~\mu$Jy and within $4.55'$ of the field center where the
sensitivity drops to 8\% of the on-axis sensitivity.  Sixteen
additional sources, not in the complete sample, are also included in
the Table.  All instrumental and resolution effects used to obtain the
source parameters have been described elsewhere \citep{fom91,win93}.

     Table 1 is organized as follows: Columns 1 and 2 show the source
numbers used in this paper and in Paper I, respectively, to identify
the sources more conveniently.  Column 3 gives the source name.  An
asterisk before the name indicates the fifteen sources which are {\bf
not} in the complete sample.  Column 4 shows the total sky flux
density and error estimate after correction for all instrumental and
resolution effects including the primary beam attenuation of the
telescopes.  Column 5 shows the signal-to-noise on the
$6''$-resolution image and column 6 gives the deconvolved source
angular size or limit in arcseconds.  The right ascension and
declination, with rms errors are given in columns 7 and 8,
respectively.  The remaining four columns list the optical
identification type and integrated magnitude in three bands.  Further
descriptions of the optical data are given in the next section.

    Eight faint radio sources, which are below our formal completeness
level but have probable identifications with optical counterparts, are
listed at the end of Table 1, and labeled $a$ through $g$.  The
probability that these radio sources is real is less than 30\% based
on the radio flux density alone.  However, their near coincidence with
an optical counterpart increases their reality to 80\%; hence, only
one of the eight sources is likely to be bogus \cite{ric99}.

       There is good agreement between the source list in Table 1 and
that in Paper I based on about one-half as much data, and at lower
resolution.  Of the 20 sources above the previous detection level of
$8.8~\mu$Jy, 16 are in the complete sample of Table 1.  The four
remaining sources from Paper I are just below our $7.5~\mu$Jy
completion limit; three of these objects are identified with galaxies.

      The source list for the Hercules field is given in Table 2.  The
resolution is $3.5''$ and the completeness limit is $35~\mu$Jy.  Ten
sources are listed; six of these are in the complete sample.  Columns
1 through 7 are similar in content to that in Table 1.  Only two of
the radio sources have been identified from an HST image which have
been described elsewhere \citep{win98}; the optical information is
given in column 8.

\section {The Optical Identifications}

       The identification of radio sources in the SA13 field is based
on two sets of optical material.
the HST-MDS WF/PC images---observed with two filters: V-band
(5420\AA) and I-band (8920\AA), respectively---were used for the
identification of radio sources near the center of the SA13 radio
field.  The calibration and processing of this optical data was
discussed in Paper I.  The identification limits are V=26.8 mag and
I=25.8 mag with FWHP resolution of $0.2''$.  For radio sources outside
the MDS field of view, identifications were made with the KPNO 4-m
telescopes with B-band (4420\AA) and I-band (8920\AA) filters.  The
limiting magnitudes here were B=26.0 and I=25.5 mag, with seeing of
about FWHM $1.5''$. The observational details are discussed elsewhere
\citep{cam99,ric02}.  The I-magnitude scale of the KPNO images was
calibrated to agree with that of the MDS, with an estimated error of
0.2 mag.  The B-magnitude scale of the KPNO images were calibrated
using the extrapolation of the MDS I- and V-magnitude scale, assuming
a power law extrapolation of the intensity of the the flat-spectrum
(blue) objects.  This extrapolation is accurate to an estimated error
of 0.4 mag.

     The precise registration of the KPNO and the MDS images with the
radio grid was determined from the alignment of the high quality
radio/optical coincidences to an accuracy of $0.2''$.  Only zero and
first order corrections to the optical image were needed.  Confidence
in the registration of the images at this accuracy is important for the
proper interpretation of the identifications.

    The integrated optical magnitudes for all identifications in the
SA13 field are given in Table 1.  The galaxy type is given in Column 9
and the B-, V- and I-magnitudes are given in columns 10, 11 and 12,
respectively.  The galaxy types are: g=galaxy of unknown morphology;
g?=uncertain identification of a very faint object; EF = empty field;
g/b=binary galaxies or closely spaced galaxies, with the magnitude
given for the objects containing the radio emission; Q=quasar:
Ell=elliptical galaxy; sp=spiral galaxy.

     In Figures 4a and 4b we show postage-stamp plots for all 42 radio
sources in the SA13 field listed in Table 1.  In contrast to Figure 3,
the green contour lines in Figures 4a and 4b show the optical I-band
distribution, while the location of the position of the radio source
is indicated by the small black ellipses in each plot.  Each field is
$8''$ on a side and centered at the position of the radio source.
Most of the radio sources are unresolved (typically $<2.5''$) and the
black ellipses indicates the 1- and 2-$\sigma$ limits for the position
of the radio centroid.  All sources have a KPNO I-band image FWHM
$1.5''$ seeing and the 22 sources nearest the field center also have
an MDS I-band image with FWHP $0.2''$ resolution. (See Paper I for a
discussion of the removal of the chromatic aberration from the
pre-refurbished HST images.)

\vskip 0.3truein

\noindent
{\bf Comments on the SA13 Identifications:}
\begin {itemize}
\item {\bf No 1. J131157+423910:} Detected only on the I-band KPNO
image with I=23.5 mag.  The optical identification is too faint to be
classified.

\item {\bf No 2. J131157+423610:} Empty field.  Possible detection of
an I=26 mag object.

\item {\bf No 3. J131203+424030:} Detected only on the I-band KPNO
image with I=24.3 mag.  The identification is too faint to be
classified.

\item {\bf No 4. J131203+423331:} Identified with an isolated galaxy, somewhat
extended in PA=$100^\circ$.

\item {\bf No 5. J131205+423851:} Empty field.

\item {\bf No 6. J131207+423945:} Empty field.

\item {\bf No 7. J131209+424217:} Identified with an isolated galaxy,
possibly extended in pa=$70^\circ$.

\item {\bf No 8. J131211+424053:} The radio emission is identified
with the southern pair of two circular galaxies separated by $3''$.
The identified galaxy has (B-I) color =3.7 mag and a size of about
$2''$.  The northern galaxy, also circular, has about the same angular
size, but is not unusually red.  Its radio flux density is $17\pm
5~\mu$Jy, but lies below the formal detection level.

\item {\bf No 9. J131213+423333:} A weak radio source, not in the
complete sample, which is identified with a bright galaxy with a faint
companion $2''$ to the north.  Another elliptically-shaped galaxy
about $4''$ south is extended towards the radio source and may be
interacting with the bright galaxy.  Because of the faintness of the
radio source, the radio-optical position offset is uncertain,

\item {\bf No 10. J131213+424129:} Identified with a bright galaxy with
an extensive, blue halo.

\item {\bf No 11. J131213+423706:} A faint smudge is present on both
the KPNO and MDS images with I$\approx 25.5$ mag.

\item {\bf No 12. J131214+423821:} A faint radio source below the
completeness level is identified with a galaxy.  Both this galaxy and
a fainter galaxy, about $3''$ NW, are elongated in the same direction.
The radio emission is extended about $3''$, about the size of the
galaxy disk.  The light profile of the galaxy is exponential,
suggesting two interacting galaxies \citep{win94}.

\item {\bf No 13. J131214+423730:} Identified with a faint blue
galaxy.  It is too faint to be classified further.

\item {\bf No 14. J131215+423702:} Identified with a bright, distorted
galaxy with z=0.322 (Paper I).  There may be a bridge of optical
emission between this galaxy and a fainter galaxy $3''$ to the west.
The galaxy shows a bulge population and a strong $r^{1/4}$ profile.
It is relatively blue in color and shows narrow emission
lines. \citep{win94}.

\item {\bf No 15. J131215+423901:} Identified with a quasar with
z=2.561 (Paper I).  The radio source is slightly extended with an
angular size of $3''$.

\item {\bf No 16. J131216+423921:} Empty field associated with the
bright radio source.  Another faint radio source, about $12''$ south
with a flux density of $6~\mu$Jy (source b in Table 1), is probably
identified with a I=24 galaxy.  The two sources are probably
unrelated.

\item {\bf No 17. J131217+423912:} Identified with the brighter of a
pair of two overlapping galaxies.  The fainter galaxy, $2.5''$ to the
NW is 1.5 mag fainter and is clearly distorted.  The velocity profile of
the galaxy is exponential \citep{win94} and the galaxy is very red.

\item {\bf No 18. J131217+423930:} Empty field.

\item {\bf No 19. J131218+423807:} A faint radio source, not in the
complete sample, is probably identified with a faint 25 mag galaxy or
galaxies.  A brighter 23-mag galaxy about $4''$ to the west appears
unrelated.

\item {\bf No 20. J131218+423843:} Identified with the brighter pair
of two nearly overlapping galaxies.  The fainter galaxy, $2.5''$ to
the SE is 1.5 mag fainter and is is more distorted than the brighter
galaxy, which is very red.

\item {\bf No 21. J131219+423608:} Identified with a distorted galaxy.
The radio source is extended $4.0''$ in position angle $120^\circ$.
From the higher resolution observations at 1.4 GHz \citep{ric02}, the
radio emission is composed of two small-diameter sources whose
positions are indicated by the two black ellipses in the image.  Since
the two radio components lie in about the same orientation as the
optical object, they are probably both associated with the galaxy.
The cause of this non-nuclear radio emission is unknown.  The radio
objects may be a small compact double source, resembling a FR-1
morphology, or they maybe associated with individual galactic objects far
from the galactic center.

\item {\bf No 22. J131219+423631:} The higher resolution observations
at 1.4 GHz suggest that this extended radio source is probably
composed of two separated radio sources \citep{ric02}.  The fainter
radio component to the NW is coincident with a bright,
elliptically-shaped galaxy.  The brighter radio component to the SE is
unidentified with a limit of I=26 mag.  We believe these two radio
components are unrelated, although the unidentified source could be
associated with a galactic component in the halo of the bright galaxy.
The light profile of the galaxy is exponential, indicating a
disk-dominated galaxy \citep{win94}.

\item {\bf No 23. J131220+423923:} Empty field.  However, this radio
sources is only $5''$ west of the bright 18-m galaxy associated with
J131221+423923 (No. 27) \citep{ric00}.  The question arises if this
radio source is associated with a galactic component in the halo of
this bright galaxy?

\item {\bf No 24. J131220+423704:} Empty field.

\item {\bf No 25. J131220+424029:} Empty field. 

\item {\bf No 26. J131220+423535:} Empty field. 

\item {\bf No 27. J131221+423923:} Identified with an 18-m barred
spiral galaxy with z=0.302 (Paper I).  The radio source is about $5''$
in size, contained within the disk of the galaxy.  The galaxy disk is
distorted; it is blue in color and there are narrow emission lines
\citep{win94}.

\item {\bf No 28. J131221+423722:} Identified with an 18-m, edge-on,
barred spiral galaxy with z=0.180 (Paper I).  The galaxy has a
relatively blue color and shows weak emission lines.  More details are
given elsewhere \cite{win94}.

\item {\bf No 29. J132121+423827:} Identified with the middle optical
peak in a chain of three or four galaxies comprising one of the most
distorted optical regions associated with a detected radio source.
The light profile of the galaxy is exponential, indicating a disk
galaxy. \citep{win94}.

\item {\bf No 30. J131222+423813:} Identified with a quasar with
z=2.561, i.e.\ the same redshift for J131315+423901 (No 15).  They
are probably in the same distant cluster or group of objects.  With a
separation of $80''$ they are unlikely to be a gravitational-lens
pair.

\item {\bf No 31. J131223+423908:} Identified with the brighter of a
pair of galaxies, separated by $6''$ .  The brighter galaxy is
somewhat distorted and the fainter galaxy is slightly extended towards
the brighter.  The radio emission is also slightly extended, although
well within the disk of the brighter galaxy.  The velocity profile of
the galaxy is exponential and its color is very red \citep{win94}.

\item {\bf No 32. J131223+423712:} Identified with the brighter of a
pair of overlapping galaxies, separated by $2''$ .  Two other
overlapping galaxies lie $5''$ to the east, and all four may be
connected by a faint optical bridge.  The radio emission is coincident
with the galactic nucleus of the bright galaxy and is $<2''$ in
angular size.  Higher resolution observations at 1.4 GHz \citep{ric02}
give a size of $1.5''$ in PA=$120^\circ$.  This radio source may be
associated with an interacting system of galaxies.  The redshift of
both bright galaxies is 0.401 (Paper I).  The galaxy containing the
radio source has little significant bulge population, contains weak
emission lines and is rather red \citep{win94}.

\item {\bf No 33. J131223+423525:} Identified with a bright, somewhat
elliptical galaxy with a large distorted halo.

\item {\bf No 34. J131224+423804:} A faint radio source, not in the
complete sample, which is identified with a faint object or
objects. However, the identification is too faint to be further
categorized.

\item {\bf No 35. J131225+424656:} Possibly identified with a I=25 mag
object, although the radio/optical offset is larger than expected.

\item {\bf No 36. J131225+424103:} A relatively bright radio source
which lies $2''$ south of the I=22 mag Ir galaxy and $4''$ north of an
I=20 mag asymmetric galaxy.  The radio source may be associated with a
faint extension south from the Ir galaxy with I=25.5 mag.  The faint
optical extensions associated with the two bright galaxies indicate
they are probably interacting with each other.

\item {\bf No 37. J131225+423941:} A bright radio source which may be
associated with an object just at the detection limit.  It is $1.5''$
north of an I=23.5 mag galaxy.  This radio source lies near the MDS
field edge.

\item {\bf No 38. J131226+424227:} Identified with an I=18 mag, slightly
elliptical galaxy.  The radio source is not in the complete
sample.

\item {\bf No 39. J131227+423800:} Identified with a I=18mag edge-on
galaxy which is $5''$ SSE of an I=17 mag elliptical galaxy (the
contours are distorted near this galaxy).  These two galaxies overlap,
although there is no appreciable distortion; hence, their interaction
is not certain.  The light profile of the galaxy is exponential and
its color is unusually red, possibly a dusty, disk galaxy
\citep{win94}.

\item {\bf No 40. J131232+424038:} Identified with an I=24.8 mag
object which may be extended in the east-west direction.  The
identification is too faint to be further classified.

\item {\bf No 41. J131236+424027:} This radio source is displaced in
PA=$270^\circ$ from the nucleus of this I=18 mag galaxy.  It is not in the
complete radio sample so the four-sigma radio/optical offset might be
spurious.  The galaxy is elliptical and symmetric with little
distortion.  However, the galaxy orientation is also in PA=$270^\circ$,
which somewhat supports the radio/optical offset.

\item {\bf No 42. J131239+423911:} Identified with an I 24.5 mag
galaxy which is too faint to be further classified.

\item {\bf Weak Source a. J131213+423946:} Identified with an I=20.0
mag circular galaxy.  Another galaxy with I=22.5 mag lies $4''$ to the
south-west.

\item {\bf Weak Source b. J131213+423826:} Identified with a faint
galaxy which is too faint to be further classified.

\item {\bf Weak Source c. J131215+423913:} Identified with a diffuse,
extended galaxy.

\item {\bf Weak Source d. J131216+423920:} Radio source is $2''$ east
of an I=23.8 mag, somewhat extended, galaxy.

\item {\bf Weak Source e. J131219+423932:}  Radio source near an I=24.0 mag galaxy.
Brighter galaxies lie $2''$ to the south-west and $3''$ to the north-east.

\item {\bf Weak Source f. J131221+423836:} Companion galaxy lies $3''$
east-north-east.

\item {\bf Weak Source g. J131225+423742:} Bright elliptical galaxy
with I=18.0 mag.  Radio source may be extended in the same direction.

\item {\bf Weak Source h. J131225+423907:} Identified with a slightly
extended I=22.5 mag galaxy.

\end{itemize}
\section {The Properties of the Micro-Jansky Sources at 8.4 GHz}

    We have gathered data from five deep VLA integrations at 8.4 GHz
in order to study the radio spectral properties, the optical
properties and the source count of the micro-Jansky radio population.
Parameters for the five surveys made over the past decade are
summarized in Table 3, which includes references to lower frequency
radio data and optical data as well.  Column 5 gives the integration
time in each field, and column 6 gives the complete survey detection level.
Column 7, 8 and 9 show the total number of sources listed in each
survey, the number of sources in the complete sample above the detection
level, and the number of sources expected based on the best source
count described below; respectively.

    Although the five catalogs contain a total of 201 sources, only 89
of them are in a complete flux-limited 8-GHz sample, and only these are
listed in Table 4.  The table contains the source properties that are
discussed in this section.  The last column indicates the detection
of X-ray emission from the sources in the Hubble and SA13 fields
\citep{bra01, mus00}

\subsection{Spectral Index}

   Accurate spectral indices of the 89 sources have been determined by
comparing our results to observations at frequencies lower than 8.4
GHz (see references at the end of Table 3).  Only seven of them are
lower limits (one in the SA13 field and six in the Hubble fields).  In
Figure 5 we show four histogram comparison of the spectral index
distributions associated with various radio and optical properties.

   In Figure 5a we show the spectral index distribution for all 89
sources and separately for the 34 sources in the SA13 field.  There is
no significant difference between the two distributions.  In the total
distribution, the spectral range of the seven sources which were not
detected above twice the 1.4-GHz noise level upper limits on the
spectral index, are indicated by the arrows in the histogram.  Since
the observations at the various frequencies were made at different
times, source variability could affect the spectral index calculation,
but variability of micro-Jansky sources is not common
\citep{oor85,ric99}.  About one-quarter of the sources have a spectral
index in the range $-0.85<\alpha<-0.65$.  The distribution drops
precipitously at lower spectral indices (steeper spectra), but more
slowly for high spectral indices (flatter spectra).  The median value
is about $\alpha_m=-0.60$.  Only a few objects have an inverted
spectral index with $\alpha>+0.1$, although some sources with only a
1.4 GHz flux density upper limit could fall in this category.

    The spectral distribution shows a higher proportion of steep
spectrum objects than reported from less sensitive, 8.4-GHz
micro-Jansky surveys where the median spectral index was
$\alpha_m\approx -0.4$ \citep{fom91,win93}.  The reason for this
difference is illustrated in Figure 5b where the spectral distribution
is shown for 46 sources with S$_{8.4}>35~\mu$Jy and for 43 weaker
sources with S$_{8.4}<35~\mu$Jy.  {\it The lower flux density sources
are more likely to have a steep spectral index than the high flux
density sources.}  The likelihood that the distribution of two samples
chosen randomly would differ by at least this much is only 18\%.
This change explains why a flatter average spectrum was found in the less
sensitive, earlier, 8.4 GHz surveys.  Possible causes of the change of
spectral properties below $35~\mu$Jy are given in $\S$6.

   Figure 5c shows that those objects which are fainter than I=25.5
mag are predominantly steep radio spectrum objects.  Although there
are only 17 sources in this faint galaxy sample; no object has a
spectral index flatter than $-0.2$ and most are steeper than $-0.65$.
Figure 5d is a plot of the spectral index distribution for the 41
galaxies which have measured colors.  As a discriminatory of the color
for the SA13 field we used the (B-I) magnitudes; for the Hubble field
we used the [I-(H+K)/2] magnitudes.  In this sample of 41 sources, 19
galaxies are significantly reddened and their spectral index
distribution is also shown.  The slight difference in the radio
spectral index between the two groups suggests that fewer of reddened
objects have a flat spectrum.  This trend was also found from VLA
observations at 5 GHz and 1.4 GHz and from the Canadian-France
redshift Survey \citep{fom91,ham95}.  This correlation, although
weaker than that in Figure 5c, is compatible since reddened objects
would tend to have relatively faint I-band emission.
   
   Nine of the sources in Table 4 are identified with X-ray sources
\citep{mus00,bra01}.  Two are in the SA13 field.  In contrast, seven
X-ray identifications have been made in the Hubble field and these
associations show no trends in radio spectral index or optical
counterpart, and are identified with galaxies of various types.

\subsection {The 8.4-GHz Source Count}

   The density of sources (source count) is generally specified as $dn
= n_0S^\gamma dS$ where $dn$ is the number of sources per solid angle
per flux density interval $dS$ at a flux density $S$.  For a flat
space which is uniformly filled with radio sources and ignoring the
attenuation effects of redshift, the slope $\gamma =-2.5$.  In order
to determine the two parameters, $n_0$ and $\gamma$, from the 89
source sample determined from five surveys with different
sensitivities, we used a modification of the maximum likelihood method
(MLM) originally developed for a uniformly obtained sample
\citep{cra70}.  For a range of test models, we chose reasonable
parameters, $n_0$ and $\gamma$, and calculated the relative
probability that this model would produce the observed flux density
distribution of the 89 sources, using the appropriate constraints for
each survey.  The best fit model was the one with the highest
probability, and the error estimate was determined by the set of
models that were at least 37\% (1/e) as likely as the best model.

  The number of micro-Jansky sources, $dn$ (arcmin$^{-2}\mu$Jy$^{-1}$), 
is given by

\begin{equation}
    dn = 0.0027\pm 0.0003~\Big(\frac{S}{40}\Big)^{(-2.11\pm 0.13)} dS
\end{equation}

The corresponding integral count, $N$ (arcmin$^{-2})$ is

\begin{equation}
    N = 0.099\pm 0.010~\Big(\frac{S}{40}\Big)^{(-1.11\pm 0.13)}
\end{equation}
where N is the number of sources above a flux density $S$ in
micro-Jansky.  The value, 40 $\mu$Jy, is the mean flux density of the
sources, and in this form, the two parameters $n_0$ and $\gamma$ are
nearly orthogonal.  Above $40~\mu$Jy the source density is 0.1
(arcmin)$^{-2}$, or $1.17\times 10^6$ sources sr$^{-1}$.  At the
minimum detection level of these surveys, $7.5~\mu$Jy, there are 0.64
sources (arcmin)$^{-2}$, or $7.5\times 10^6$ sources sr$^{-1}$.
Extrapolation to weaker flux density with the same slope gives 6
sources (arcmin)$^{-2}$ above $1~\mu$Jy, corresponding to an average
source separation of $25''$.

    The density of sources is still far from the natural confusion
limit when sources begin to blend with each other.  If $C_\beta$ is
the proportion of sky covered with emission, significant blending
occurs for $C_\beta > 0.05$.  If we assume an average source size of
$2''$. then then $C_{7.5}=0.0007$.  Extrapolating to $1~\mu$Jy and
$0.1~\mu$Jy gives $C_{1.0}=0.006$ and $C_{0.1}=0.08$, respectively.
Unless the average angular size of most sources is less than $1''$,
radio source blending will become a problem below about $0.2~\mu$Jy.
Recent results from MERLIN, however, shows that some faint radio
sources contain a significant part of their emission within an angular
size of $0.5''$ \citep{mux01, fom02}.

    The plot of the {\it integral} source count at 8.4 GHz is shown in
Figure 6.  The ordinate scale has been normalized to the Euclidean
integral slope of $\gamma +1=-1.5$.  At any point in the plot, the
ordinate value times S$^{-1.5}$ gives the number of sources
(arcmin)$^{-2}$ with a flux density greater than $S~\mu$Jy.  The count
derived from the MLM calculation, describe above, is shown by the
solid line.  This line extends to $1~\mu$Jy, although we cannot
exclude that the slope of the counts change below our detection limit
of $7.5~\mu$Jy.  The plotted points at the micro-Jansky levels have
been obtained by gridding the data in Table 4 into five flux density
bins.  The results of this binned data are shown in Table 5.  The MLM
fit is clearly compatible with the gridded data, as it should be.  The
number of sources in each of the five bins listed in column 4 agrees
reasonably well with the number predicted by the MLM fit, given in
column 5.

We have added in the count of sources at 8.4 GHz at high flux
densities \citep{win93}.  There are no extensive surveys of sources at
8.4 GHz of intermediate flux density sources between one and eight
milli-Janskys.  The source count at 8.4 GHz derived at micro-Jansky
levels is consistent with that at the milli-Jansky and Jansky levels.
Extrapolation of the micro-Jansky count falls slightly under the
milli-Jansky points, the excess being produced by the rapid evolution
of the brightest radio sources in the sky.

    The dotted line shows the best fit count obtained from the Lynx
and Cepheus fields only \citep{win93} which contained only 20 sources
to a detection level of $14.4~\mu$Jy.  The `old' count and the `new'
count differ at the 1.4-$\sigma$ level near $40~\mu$Jy.  There may be
several reasons for the discrepancy: First, the difference may be real
and reflect the anisotropic distribution of sources in the sky on
scales of tens of arcmin.  Such differences between the number of
sources among radio fields have already been noted by others at
various frequencies (Windhorst et al.\ 1985; Windhorst et al.\ 1993;
Hopkins et al.\ 1998; Hopkins et al.\ 1999; Richards 2000).  Second,
the optical and radio selection criteria for the various fields were
not the same.  The Hubble and SA13 fields were deliberately selected
to be radio quiet above 0.5 mJy, whereas the Hercules field contained
a known radio source of about 10 mJy.  The optical constraints for the
HDF selection (no relatively bright objects) may also be relevant.
However, since most of the radio sources and statistics come from
sources which are in the 10 to 200 $\mu$Jy flux density range at 8
GHz, the systematic effect of the avoidance of radio fields with
sources stronger than 500 $\mu$Jy is unclear.  Third, the Lynx and
Cepheus fields were observed with lower resolution than the Hubble and
SA13 fields.  Although the effects of source blending were considered
for the Lynx and Cepheus fields \citep{win93}, the blending may be
more important than originally thought.  The average angular size of
the sources in these two fields were also larger than in other fields,
and perhaps a consequence of the blending of weak sources.

    Regardless of the reasons for the somewhat different count of
sources in the Lynx and Cepheus fields, the new results are based on
more than four times the number of sources found in five different
areas and with higher resolution.  When extrapolated to $1~\mu$Jy, the
new counts give a density of $6\pm 3$ (arcmin)$^{-2}$, compared with
the previous value of $20 \pm 8 $ (arcmin)$^{-2}$.  This difference is
about 1.8-$\sigma$ and should not be regarded as significant.

    The average sky brightness at 8.4 GHz, contributed by sources with
flux densities above $7.5~\mu$Jy is $2\times 10^{-4}$K.  Extrapolating
to weaker sources, the sky brightness will increase as S$^{\gamma
+2}$, assuming a constant slope $\gamma$.  Using $\gamma =-2.11$, the
estimated sky brightness from discrete sources increases by only a
factor of 2 for a decrease of 1000 in flux density.  Hence, the sky
brightness at 8.4 GHz contributed by discrete sources down to the
nano-Jansky level is unlikely to exceed 0.001 K unless there is a new
population of weaker sources, allowing the slope $\gamma$ to become
more negative (i.e.\ steeper) than $-2.2$.

\subsection {Identifications in the SA13 and Hubble Fields}

    Figure 7 shows the distribution of the I-magnitudes of the 63
sources in the complete samples in the SA13 field plus the Hubble Deep
and Flanking fields.  The density of sources with radio
identifications and optical magnitudes between I=17.5 and 20.5 mag is
about 30\% of the density of all galaxies within this magnitude range
\citep{bur94}.  For galaxies fainter than I=20.5 mag the distribution
remains relatively constant per magnitude interval to I=26.0 mag, with
about 20\% of the sources unidentified below this magnitude.  This
distribution is similar to that found in other investigations of the
identification of weak radio sources \citep{ham95, kel99}.

    Analysis of several HST fields shows that 34\%$\pm$9\% of all
galaxies brighter than I=23.3 mag are in binary systems.  This is much
higher than the 7\% found for the galaxy population brighter than I=15
mag, which comprise the \lq local\rq~population \citep{bur94}.  Thus,
galaxy interaction appears to be a strong function of cosmological
time, and such galaxy interactions may be the main cause of the strong
evolution of radio sources with redshift \citep{wad01}.

    This connection between galaxy interaction and radio emission is
supported in these data.  For the complete sample of 63 radio sources
in the SA13 and Hubble fields, 37 are identified with galaxies
brighter than I=23.3 mag.  Seventeen of these identifications are
classified as {\it g/b} meaning that at least one other galaxy lies
within $5''$ of the radio galaxy, possibly indicating galaxy
interactions.  Thus, 46\% of these radio identifications are
associated with multiple systems.  This percentage is somewhat larger
than the 34\% of galaxies that are found in multiple systems below
$\approx 18$ mag, and suggests that the radio emission is even more
snhanced by galaxy interactions.

     A more complete analysis of the optical properties of the radio
sources in the SA13 and Hubble fields is given elsewhere
\citep{ric00,ric02,fom02}.

\section {Discussion and Conclusions}

     We have reported the results from two new 8.4 GHz VLA surveys of
micro-Jansky radio sources.  In one of the fields, SA13, the rms noise
is $1.49~\mu$Jy and we have detected 49 sources of which 34 are in a
complete sample with a flux density above $7.5~\mu$Jy.  Six sources
were detected above $35~\mu$Jy in the Hercules field.  In Tables 1 and
2 we present the radio catalog for these two fields.  Comments and a
detailed radio/optical comparison are given for 42 sources in the SA13
field using deep HST and KPNO images.  The results from these two
fields double the number of 8.4-GHz radio sources now known at
micro-Jansky levels.

     In the second part of this paper, we have compiled a list of 89
sources in a flux density complete sample from five deep VLA surveys
at 8.4 GHz.  This source list is dominated by observations of the SA13
field (detection level of $7.5~\mu$Jy) and the Hubble deep field plus
the flanking fields (detection level of $9.0~\mu$Jy).  We have used
other optical and radio data in order to determine the properties of
these weak radio sources.

\subsection {The Density and Angular Size of the Micro-Jansky Sources}

     The count of radio sources follows an approximate power-law with
a slope $\gamma=-2.11$ from 10 mJy to less than $10~\mu$Jy, with good
continuity to the count at the Jansky level (see Figure 6).  There are
no significant differences in the number of sources detected in each
field, as shown by the comparison of columns 8 and 9 in Table 3, which
indicate the number of sources detected versus the number expected
from the best fit to the count.  The average sky brightness
temperature contributed by the micro- and nano-Jansky radio sources is
extrapolated to be less than 0.001 K as long as the slope $\gamma$
does not become steeper than $-2.2$.

   Most of the sources are unresolved with $6''$ or even $4''$
resolution.  For the 63 sources in the Hubble and SA13 samples, 15
sources (24\%) are larger than about $3''$, and only four sources are
larger than $5''$.  Thus, the angular size of the micro-Jansky sources
at 8.4 GHz appears statistically somewhat smaller than the angular
size at 1.4 GHz \citep{ric00}.  The median spectral index of these 15
extended sources is $\alpha_m=-0.81$, steeper than than our general
sample and consistent with an increasing angular size with lower
frequency.  The correlation of large angular size with steeper
spectral index is seen for all radio sources, regardless of their flux
density.  The combination of emission from opaque small-diameter
components and large components showing an aging synchrotron (steep)
spectrum probably cause this difference in the correlation of angular
size and spectral index at the two different frequencies.  Higher
resolution observations are needed in order to resolve the majority of
the micro-Jansky radio sources \citep{mux01,fom02}.

\subsection {The Radio Identifications}

     The radio sources are identified with a variety of galaxy types.
Only two of the radio sources are identified with quasars and none
with stars.  We estimate that about 30\% of all galaxies brighter than
I=21 mag can be detected at the $10~\mu$Jy level at 8.4 GHz.  For
fainter magnitudes, the radio source identifications are spread
relatively evenly, per magnitude interval, with about 20\% of the
radio sample fainter than I=26 mag.  The centroid of the radio
emission is nearly always located within $0.5''$ of the galaxy
nucleus.  The few exceptions in the SA13 field are: J131220+423923 is
an empty field which is displaced $5''$ from an I=18-mag galaxy;
J131219+423608 contains two radio components each displaced about
$2''$ on opposite sides of a I=22 mag galaxy nucleus and is possibly a
small double radio source (see notes); J131225+424103 is displaced
$2''$ south of an I=22 mag galaxy (see notes); J131239+423911 is
displaced $1''$ from the galaxy centroid, but well within the disk
component.

     The likelihood that a micro-Jansky radio source will be associated
with a galaxy, between I=18 to 22 mag with at least one neighbor
closer than $5''$, is nearly 50\%.  This percentage is somewhat
larger than the 34\% of all such galaxies which are in groups.  This
suggests that radio source luminosity and/or density evolution is
enhanced somewhat within galaxy groups which may be interacting.

\subsection {The Radio Spectrum}

     The spectral index distribution of the micro-Jansky radio sources
at 8.4 GHz shows one clear trend; the spectral index distribution
steepens for sources below $35~\mu$Jy (see Figure 5b).  Somewhat less
conclusively, the spectral index distribution may also steepen for
sources with optical counterparts fainter than I=25.5 mag (see Figure
5c).

     The change in the source count and spectral properties of radio
sources fainter than 1 mJy is well-documented.  Below this level,
star-forming galaxies begin to dominate the source population over AGN.
The cause for a further change in spectral properties less than
$35~\mu$Jy may be associated with the different evolution of AGN and
star-burst galaxies, but a definitive answer must await for additional
observations of the higher energy photons emitted from these fainter
radio sources, and measurement of their radio size and internal
complexity.  Both the AGN and the star-burst phenomena produce a wide
range in spectral index, and cannot be used as a good discriminator
between the two types.  A steepening of the spectral index below
$35~\mu$Jy could arise in several ways: (1) The opaque emission
(always with flat spectrum) from AGN components in weak sources may be
a smaller proportion of the total emission; (2) The star-burst
phenomenon may produce steeper radio spectra for the weak sources,
possibly because of synchrotron aging associated with larger magnetic
fields in the weaker, distant sources.

\acknowledgements

The National Radio Astronomy Observatory is a facility of the National
Science Foundation, operated under cooperative agreement by Associated
Universities, Inc.  The Space Telescopes Science Institute is operated
by Associated Universities for Research in Astronomy, Inc., under
contract to NASA.  RBP was supported in part by NSF grant AST96-16971
and by the Keck Northeast Astronomy Consortium.  EAR was supported
by a Hubble Fellowship.

\begin{deluxetable}{rrrrrrcrcrrr}
\rotate
\tablecolumns{12}
\tablewidth{0pt}
\tabletypesize{\scriptsize}
\tablecaption{SA13 SOURCE LIST AND IDENTIFICATIONS}

\tablehead{
 \multicolumn{2}{c} {Num} &
 \colhead {Source} &
 \colhead {S$_{8.4}$} &
 \colhead {SNR} &
 \colhead {Size} &
 \colhead {RA (J2000)} &
 \colhead {DEC (J2000)} &
 \multicolumn{4}{c}{Identification} \\

 \colhead {New} &
 \colhead {Old} &
 \colhead {} &
 \colhead {($\mu$Jy)} &
 \colhead { } &
 \colhead {($''$)} &
 \multicolumn{2}{c}{} &
 \colhead {Type} &
 \colhead {Bmag} &
 \colhead {Vmag} &
 \colhead {Imag} \\
}
\startdata

  1 &    &  J131157$+$423910 & 109.3 $\pm$  8.1 &  14.8 & $<2.0$ & 13 11 57.499  $\pm$  0.02 & 42 39 10.16  $\pm$ 0.21 & g   & $>$26.0 & $ $     & $  $ 23.5 \\
  2 &    &  J131157$+$423630 & 196.6 $\pm$ 10.2 &  23.2 & $<2.0$ & 13 11 57.903  $\pm$  0.01 & 42 36 30.15  $\pm$ 0.14 & EF  & $>$26.0 & $ $     & $> $ 25.5 \\
  3 &    & *J131203$+$424030 & 213.2 $\pm$  8.7 &  36.7 & $<2.0$ & 13 12 03.092  $\pm$  0.01 & 42 40 30.68  $\pm$ 0.10 & g   & $ $25.5 & $ $     & $  $ 24.3 \\
  4 &    &  J131203$+$423331 & 886.1 $\pm$ 60.9 &  15.6 & $<2.0$ & 13 12 03.508  $\pm$  0.02 & 42 33 31.72  $\pm$ 0.20 & g   & $ $24.0 & $ $     & $  $ 22.2 \\
  5 &    &  J131205$+$423851 &  12.4 $\pm$  2.6 &   5.0 & $<3.0$ & 13 12 05.596  $\pm$  0.06 & 42 38 51.15  $\pm$ 0.60 & EF  & $>$26.0 & $ $     & $> $ 25.5 \\
  6 &    &  J131207$+$423945 &  14.0 $\pm$  2.8 &   5.4 & $<2.5$ & 13 12 07.645  $\pm$  0.06 & 42 39 45.87  $\pm$ 0.56 & EF  & $>$26.0 & $ $     & $> $ 25.5 \\
  7 &    &  J131209$+$424217 &  93.4 $\pm$ 17.8 &   5.8 & $<2.5$ & 13 12 09.068  $\pm$  0.05 & 42 42 17.49  $\pm$ 0.52 & g   & $ $25.5 & $ $     & $  $ 23.4 \\
  8 &    &  J131211$+$424053 & 717.0 $\pm$ 21.9 & 183.0 & $<2.0$ & 13 12 11.029  $\pm$  0.01 & 42 40 53.66  $\pm$ 0.05 & g/b & $ $24.5 & $ $     & $  $ 20.8 \\
  9 &    & *J131213$+$423555 &  12.2 $\pm$  2.6 &   4.7 & $<3.0$ & 13 12 13.303  $\pm$  0.07 & 42 35 55.76  $\pm$ 0.64 & g/b & $ $21.7 & $ $     & $  $ 19.5 \\
 10 &    &  J131213$+$424129 &  48.4 $\pm$  5.6 &   9.2 & $<2.5$ & 13 12 13.446  $\pm$  0.04 & 42 41 29.44  $\pm$ 0.33 & ell & $ $21.9 & $ $     & $  $ 18.7 \\
 11 &    &  J131213$+$423706 &   9.4 $\pm$  1.7 &   5.6 & $<2.5$ & 13 12 13.858  $\pm$  0.06 & 42 37 06.50  $\pm$ 0.53 & g?  & $ $     & $>$26.8 & $  $ 25.5 \\
 12 &  1 & *J131214$+$423821 &   6.5 $\pm$  1.6 &   4.3 & $ 3.5$ & 13 12 14.518  $\pm$  0.08 & 42 38 21.86  $\pm$ 1.02 & g/b & $ $23.7 & $ $23.4 & $  $ 21.8 \\
 13 &  2 & *J131214$+$423730 &   6.2 $\pm$  1.6 &   4.4 & $<3.0$ & 13 12 14.666  $\pm$  0.09 & 42 37 30.96  $\pm$ 1.13 & g   & $ $     & $ $26.3 & $  $ 25.2 \\
 14 &  3 &  J131215$+$423702 &  13.4 $\pm$  1.7 &   7.8 & $<2.5$ & 13 12 15.130  $\pm$  0.04 & 42 37 02.70  $\pm$ 0.39 & g/b & $ $21.4 & $ $20.7 & $  $ 19.7 \\
 15 &  4 &  J131215$+$423901 &  26.6 $\pm$  1.8 &  14.9 & $ 3.0$ & 13 12 15.280  $\pm$  0.02 & 42 39 01.16  $\pm$ 0.21 & Q   & $ $18.6 & $ $18.4 & $  $ 17.8 \\
 16 &  5 &  J131216$+$423921 &  29.5 $\pm$  2.0 &  15.4 & $<2.5$ & 13 12 16.083  $\pm$  0.02 & 42 39 21.47  $\pm$ 0.20 & EF  & $ $     & $>$26.8 & $> $ 25.8 \\
 17 &  6 &  J131217$+$423912 &  20.1 $\pm$  1.8 &  11.7 & $<2.0$ & 13 12 17.181  $\pm$  0.03 & 42 39 12.15  $\pm$ 0.26 & g/b & $ $22.5 & $ $22.2 & $  $ 20.0 \\
 18 &  7 &  J131217$+$423930 &  15.5 $\pm$  1.9 &   8.6 & $<2.5$ & 13 12 17.601  $\pm$  0.04 & 42 39 30.49  $\pm$ 0.35 & EF  & $ $     & $>$26.8 & $> $ 25.8 \\
 19 &  8 & *J131218$+$433907 &   6.4 $\pm$  1.6 &   4.3 & $<3.0$ & 13 12 18.311  $\pm$  0.09 & 42 39 07.79  $\pm$ 1.07 & g?  & $ $     & $>$26.8 & $  $ 25.2 \\
 20 &  9 &  J131218$+$423843 &  25.0 $\pm$  1.7 &  14.8 & $ 5.0$ & 13 12 18.437  $\pm$  0.02 & 42 38 43.92  $\pm$ 0.21 & g/b & $ $24.0 & $ $23.1 & $  $ 20.3 \\
 21 &    &  J131219$+$423608 &  26.4 $\pm$  2.4 &  11.4 & $ 4.0$ & 13 12 19.838  $\pm$  0.03 & 42 36 08.91  $\pm$ 0.27 & g   & $ $23.8 & $ $     & $  $ 22.1 \\
 22 & 10 &  J131219$+$423831 &  11.6 $\pm$  1.6 &   6.7 & $ 4.5$ & 13 12 19.932  $\pm$  0.05 & 42 38 31.39  $\pm$ 0.45 & g   & $ $23.0 & $ $22.2 & $  $ 20.4 \\
 23 & 12 &  J131220$+$423923 &  14.5 $\pm$  1.9 &   7.7 & $<2.5$ & 13 12 20.016  $\pm$  0.04 & 42 39 23.97  $\pm$ 0.39 & EF? & $ $     & $>$26.8 & $> $ 25.8 \\
 24 & 11 &  J131220$+$423704 &  13.9 $\pm$  1.7 &   8.4 & $<2.5$ & 13 12 20.185  $\pm$  0.04 & 42 37 04.02  $\pm$ 0.36 & EF  & $ $     & $>$26.8 & $> $ 25.8 \\
 25 &    &  J131220$+$424029 &  33.7 $\pm$  3.0 &  11.5 & $<2.0$ & 13 12 20.236  $\pm$  0.03 & 42 40 29.67  $\pm$ 0.26 & EF  & $>$26.0 & $ $     & $> $ 25.5 \\
 26 &    &  J131220$+$423535 &  20.4 $\pm$  3.0 &   6.7 & $<2.5$ & 13 12 20.871  $\pm$  0.05 & 42 35 35.33  $\pm$ 0.45 & EF  & $>$26.0 & $ $     & $> $ 25.5 \\
 27 & 13 &  J131221$+$423923 &  11.5 $\pm$  1.9 &   5.4 & $ 5.0$ & 13 12 21.109  $\pm$  0.06 & 42 39 23.49  $\pm$ 0.55 & sp  & $ $20.2 & $ $19.2 & $  $ 18.0 \\
 28 & 15 &  J131221$+$423722 &  10.0 $\pm$  1.7 &   5.9 & $<2.5$ & 13 12 21.397  $\pm$  0.05 & 42 37 22.91  $\pm$ 0.51 & sp  & $ $20.0 & $ $19.0 & $  $ 17.8 \\
 29 & 16 &  J131221$+$423827 &  18.5 $\pm$  1.7 &  10.2 & $<2.5$ & 13 12 21.839  $\pm$  0.03 & 42 38 27.55  $\pm$ 0.30 & g/b & $ $     & $ $24.3 & $  $ 22.7 \\
 30 & 17 &  J131222$+$423813 &  10.4 $\pm$  1.6 &   5.6 & $<3.5$ & 13 12 22.428  $\pm$  0.06 & 42 38 13.47  $\pm$ 0.54 & Q   & $ $21.3 & $ $21.0 & $  $ 19.6 \\
 31 & 18 &  J131223$+$423908 &  19.3 $\pm$  2.0 &   9.2 & $ 2.5$ & 13 12 23.298  $\pm$  0.04 & 42 39 08.35  $\pm$ 0.33 & g/b & $ $22.6 & $ $21.6 & $  $ 19.3 \\
 32 & 19 &  J131223$+$423712 &  27.7 $\pm$  2.0 &  15.2 & $<2.0$ & 13 12 23.693  $\pm$  0.02 & 42 37 12.09  $\pm$ 0.20 & g/b & $ $21.3 & $ $20.6 & $  $ 18.8 \\
 33 &    &  J131223$+$423525 &  65.5 $\pm$  4.1 &  17.7 & $<2.0$ & 13 12 23.999  $\pm$  0.02 & 42 35 25.57  $\pm$ 0.18 & Ell & $ $22.4 & $ $     & $  $ 18.9 \\
 34 & 20 & *J131224$+$433804 &   6.2 $\pm$  1.6 &   4.2 & $<3.0$ & 13 12 24.133  $\pm$  0.09 & 42 38 04.89  $\pm$ 1.02 & g   & $ $     & $ $25.5 & $  $ 24.0 \\
 35 &    &  J131225$+$423656 &  12.9 $\pm$  2.1 &   6.5 & $<2.5$ & 13 12 25.201  $\pm$  0.05 & 42 36 56.66  $\pm$ 0.47 & g?  & $>$26.8 & $ $     & $  $ 25.0 \\
 36 &    &  J131225$+$424103 &  34.9 $\pm$  4.8 &   6.7 & $<2.5$ & 13 12 25.228  $\pm$  0.05 & 42 41 03.44  $\pm$ 0.45 & g?  & $>$26.0 & $ $     & $  $ 25.5 \\
 37 & 23 &  J131225$+$423941 & 200.3 $\pm$  6.5 &  79.3 & $<2.0$ & 13 12 25.743  $\pm$  0.01 & 42 39 41.58  $\pm$ 0.06 & g/b & $>$26.5 & $ $     & $  $ 25.5 \\
 38 &    & *J131226$+$424227 & 318.3 $\pm$ 25.3 &  13.0 & $<2.0$ & 13 12 26.287  $\pm$  0.03 & 42 42 27.43  $\pm$ 0.24 & Ell & $ $20.9 & $ $     & $  $ 17.7 \\
 39 & 24 &  J131227$+$423800 &  25.6 $\pm$  2.3 &  11.6 & $<2.0$ & 13 12 27.532  $\pm$  0.03 & 42 38 00.23  $\pm$ 0.26 & g/b & $ $22.1 & $ $22.0 & $  $ 19.8 \\
 40 &    &  J131232$+$424038 & 165.7 $\pm$  8.9 &  20.5 & $<3.0$ & 13 12 32.797  $\pm$  0.02 & 42 40 38.42  $\pm$ 0.15 & g   & $>$26.0 & $ $     & $  $ 24.8 \\
 41 &    & *J131236$+$424027 &  54.6 $\pm$ 11.5 &   4.8 & $<2.0$ & 13 12 36.044  $\pm$  0.07 & 42 40 27.66  $\pm$ 0.63 & g   & $ $20.2 & $ $     & $  $ 18.3 \\
 42 &    &  J131239$+$423911 &  66.2 $\pm$ 12.7 &   5.1 & $<2.0$ & 13 12 39.705  $\pm$  0.06 & 42 39 11.32  $\pm$ 0.60 & g   & $>$25.0 & $ $     & $  $ 24.5 \\
  a &    & *J131213$+$423932 &   4.2 $\pm$  1.7 &   2.5 & $<4.0$ & 13 12 13.521  $\pm$  0.26 & 42 39 32.16  $\pm$ 1.33 & g   & $ $     & $ $     & $  $ 20.0 \\
  b &    & *J131213$+$423826 &   4.8 $\pm$  1.6 &   2.7 & $<4.0$ & 13 12 13.583  $\pm$  0.22 & 42 38 26.68  $\pm$ 1.13 & g   & $ $     & $ $     & $  $ 24.5 \\
  c &    & *J131215$+$423913 &   6.7 $\pm$  1.6 &   3.8 & $<3.0$ & 13 12 15.847  $\pm$  0.12 & 42 39 13.22  $\pm$ 0.70 & g   & $ $     & $ $     & $  $ 23.2 \\
  d &    & *J131216$+$423920 &   4.8 $\pm$  1.6 &   2.8 & $<6.0$ & 13 12 16.521  $\pm$  0.25 & 42 39 20.40  $\pm$ 2.20 & g   & $ $     & $ $     & $  $ 23.8 \\
  e &    & *J131219$+$423932 &   5.0 $\pm$  1.9 &   2.9 & $<6.0$ & 13 12 19.092  $\pm$  0.22 & 42 39 32.20  $\pm$ 1.80 & g/b & $ $     & $ $     & $  $ 24.0 \\
  f &    & *J131221$+$423836 &   4.8 $\pm$  1.8 &   2.8 & $<3.5$ & 13 12 21.794  $\pm$  0.13 & 42 38 36.96  $\pm$ 1.40 & g   & $ $     & $ $     & $  $ 22.4 \\
  g &    & *J131225$+$423742 &   5.5 $\pm$  1.8 &   3.3 & $ 6.0$ & 13 12 25.065  $\pm$  0.17 & 42 37 42.33  $\pm$ 1.50 & g   & $ $     & $ $     & $  $ 18.0 \\
  h &    & *J131225$+$423907 &   5.0 $\pm$  1.8 &   3.0 & $<4.0$ & 13 12 25.549  $\pm$  0.15 & 42 39 07.01  $\pm$ 1.60 & g   & $ $     & $ $     & $  $ 22.5 \\
\tablecomments { 
{\bf * = Not in complete sample.} {\protect \\}
{\bf Galaxy Types:}
{\bf g} - galaxy of unknown type.
{\bf g/b} - interacting galaxies.  Magnitudes given for galaxy associated with the radio source.
{\bf Q} - Quasar.
{\bf g?} - Uncertain identification with a faint object.
{\bf EF} - Empty field with magnitude limit given.
{\bf ell} - Elliptical galaxy.
{\bf sp} - Spiral galaxy.
}	
\enddata
\end{deluxetable}

\begin{deluxetable}{rrrrrrcl}
\rotate
\tablecolumns{8}
\tablewidth{0pt}
\tabletypesize{\scriptsize}
\tablecaption{HERCULES SOURCE LIST AND IDENTIFICATIONS}
\tablehead{
 \colhead {Num} &
 \colhead {Source} &
 \colhead {S$_{8.4}$} &
 \colhead {SNR} &
 \colhead {Size} &
 \colhead {RA.(J2000)} &
 \colhead {DEC (J2000)} &
 \colhead {Comments} \\

 \multicolumn{2}{c}{} &
 \colhead {($\mu$Jy)} &
 \colhead { } &
 \colhead {($''$)} \\
}
\startdata

  1 &  J171349$+$501610 &  1136 $\pm$  63 &  20.3 & $<2.0$ & 17 13 49.077  $\pm$  0.02 & 50 16 10.49  $\pm$ 0.16 \\
  2 &  J171354$+$501547 &   153 $\pm$  22 &   6.7 & $<2.5$ & 17 13 54.160  $\pm$  0.05 & 50 15 47.70  $\pm$ 0.45 \\
  3 & *J171407$+$501547 &    41 $\pm$   8 &   4.9 & $<3.0$ & 17 14 07.032  $\pm$  0.06 & 50 15 47.76  $\pm$ 0.61 \\
  4 &  J171411$+$501602 &    37 $\pm$   7 &   5.0 & $<3.0$ & 17 14 11.951  $\pm$  0.06 & 50 16 02.42  $\pm$ 0.60 & Obj 18, 23.5m g/b \\
  5 &  J171414$+$501530 &  6482 $\pm$  20 &   0.9 & $<1.5$ & 17 14 14.754  $\pm$  0.00 & 50 15 30.46  $\pm$ 0.05 & 53W002, 22.1m g/b \\
  6 &  J171415$+$501535 &    49 $\pm$   7 &   8.5 & $<3.0$ & 17 14 15.219  $\pm$  0.01 & 50 15 35.45  $\pm$ 0.15 \\
  7 &  J171416$+$501817 &   495 $\pm$  22 &  30.4 & $<2.0$ & 17 14 16.771  $\pm$  0.01 & 50 18 17.10  $\pm$ 0.11 \\
  8 & *J171424$+$501339 &    62 $\pm$  13 &   4.7 & $<3.0$ & 17 14 24.613  $\pm$  0.07 & 50 13 39.78  $\pm$ 0.64 \\
  9 & *J171436$+$501329 &   194 $\pm$  42 &   4.5 & $<3.0$ & 17 14 36.222  $\pm$  0.07 & 50 13 29.56  $\pm$ 0.67 \\
 10 & *J171442$+$501640 &   505 $\pm$  99 &   4.9 & $<3.0$ & 17 14 42.629  $\pm$  0.07 & 50 16 40.09  $\pm$ 0.62 \\

\tablecomments {
{\bf * = Not in complete sample}
}
\enddata
\end{deluxetable}

\begin{deluxetable}{llllrrrcrl}
\tablecolumns{10}
\tablewidth{0pt}
\tabletypesize{\scriptsize}
\tablecaption{PARAMETERS FOR FIVE VLA 8.4-GHZ DEEP INTEGRATIONS}
\tablehead{
 \colhead {Field} &
 \colhead {Date} &
 \colhead {RA (J2000)} &
 \colhead {DEC (J2000)} &
 \colhead {$\tau$} &
 \colhead {S$_{det}$} &
 \colhead {N$_{tot}$} &
 \colhead {N$_{cs}$} &
 \colhead {N$_{fit}$} &
 \colhead {Ref.}  \\

 \colhead {} &
 \colhead {} &
 \multicolumn{2}{c}{} &
 \colhead {(hr)} &
 \colhead {($\mu$Jy)} \\

}
\startdata

Cepheus  & 1988-89 & 03 10 00 & 80 00 00 &  30 & 20.3 & 36 &  6 & 10.2 & 1,2,3 \\
Lynx     & 1989-90 & 08 41 40 & 44 45 00 &  63 & 12.8 & 46 & 14 & 17.1 & 1,4 \\
Hercules & 1996    & 17 14 15 & 50 15 30 &  12 & 35.0 & 10 &  6 &  5.5 & 0 \\
SA13     & 1994-95 & 13 12 17 & 42 38 05 & 159 &  7.5 & 50 & 34 & 30.9 & 0,5,6 \\
Hubble   & 1996-97 & 12 36 49 & 62 12 58 & 139 &  9.0 & 60 & 29 & 25.2 & 7,8,9 \\

\tablecomments { 
{\bf References} {\protect \\}
0=This paper; 1=Windhorst et al.\ 1993; 2=Donnelly et al.\ 1987;
3=Windhorst et al.\ 1985; 4=Windhorst et al.\ 1984; 5=Windhorst et al.\ 1995;
6=Richards et al.\ 2002; 7=Fomalont et al.\ 1997; 8=Richards et al.\ 1998;
9=Richards et al.\ 2000
}	
\enddata
\end{deluxetable}

\begin{deluxetable}{lrrlrc}
\tablecolumns{6}
\tablewidth{0pt}
\tabletypesize{\footnotesize}
\tablecaption{Catalog of Micro-Jansky Sources at 8.4 GHz}
\tablehead{
 \colhead {Source} &
 \colhead {S$_{8.4}$} &
 \colhead {$\alpha$} &
 \colhead {ID} &
 \colhead {I-mag} &
 \colhead {X-ray} \\

 \colhead { } &
 \colhead {($\mu$Jy)} &
 \colhead { } &
 \colhead {Type} \\
}
\startdata
J030755+801008 & 1221 & $ -$1.0 &      & $ $     \\
J030852+801409 &  449 & $ +$0.2 &      & $ $     \\
J030902+800955 &   54 & $ +$0.2 &      & $ $     \\
J030924+801159 &   34 & $ -$0.4 &      & $ $     \\
J031005+800824 &   52 & $ -$0.2 &      & $ $     \\
J031101+800943 &  293 & $ -$0.6 &      & $ $     \\
J084120+444453 &   58 & $ -$0.5 &      & $ $     \\
J084121+444318 &  289 & $ -$0.7 &      & $ $     \\
J084126+444648 &   62 & $ -$0.6 &      & $ $     \\
J084133+444544 &   20 & $ -$0.7 &      & $ $     \\
J084134+444409 &   50 & $ -$0.2 &      & $ $     \\
J084135+444442 &   21 & $ -$0.8 &      & $ $     \\
J084136+444443 &   32 & $ -$0.6 &      & $ $     \\
J084141+444509 &   32 & $ -$0.6 &      & $ $     \\
J084141+444452 &   35 & $ -$1.0 &      & $ $     \\
J084142+444613 &  110 & $ -$0.6 &      & $ $     \\
J084142+444718 &  115 & $ -$0.2 &      & $ $     \\
J084144+444534 &   62 & $ -$0.8 &      & $ $     \\
J084150+444545 &   31 & $ -$1.1 &      & $ $     \\
J084155+444640 &   78 & $ -$0.9 &      & $ $     \\
J123632+621105 &   22 & $>-$0.1 & sp   & $ $20.1 \\
J123634+621212 &   56 & $ -$0.7 & g/b  & $ $19.3 \\
J123634+621240 &   53 & $ -$0.7 & g/b  & $ $23.5 \\
J123637+621135 &   18 & $>-$0.2 & sp   & $ $18.2 \\
J123640+621010 &   29 & $ -$0.4 & g    & $ $25.0 \\
J123641+621142 &   19 & $ -$0.3 & g/b  & $ $22.7 \\
J123642+621331 &   80 & $ -$0.9 & g?   & $ $25.8 & YES \\
J123642+621545 &   54 & $ -$0.5 & g    & $ $22.0 \\
J123644+621249 &   10 & $>-$0.4 & g/b  & $ $21.9 & YES \\
J123644+621133 & *599 & $ -$0.3 & Ell  & $ $20.5 & YES \\
J123646+621448 & *~25 & $ -$0.8 & g/b  & $ $26.0 \\
J123646+621445 &   13 & $ -$1.0 & g    & $ $23.9 \\
J123646+621404 &  190 & $  $0.0 & sp   & $ $20.0 & YES \\
J123649+621313 &   22 & $ -$0.7 & g/b  & $ $21.8 & YES \\
J123651+621030 &   26 & $ -$0.7 & sp   & $ $20.7 \\
J123651+621221 &   17 & $ -$0.7 & g    & $ $27.8 & YES \\
J123652+621444 &  185 & $  $0.1 & Ell  & $ $19.4 \\
J123653+621139 &   15 & $ -$0.8 & sp   & $ $22.3 \\
J123655+621311 &   12 & $>-$0.2 & g/b  & $ $23.1 & YES \\
J123657+621455 &   15 & $>-$0.2 & g    & $ $22.8 \\
J123700+620908 & *~67 & $ -$0.9 & g    & $ $26.3 \\
J123701+621146 & *~30 & $ -$0.7 & g/b  & $ $25.4 \\
J123707+611408 & *~29 & $ -$0.3 & g    & $ $26.4 \\
J123708+621056 & *~26 & $ -$0.4 & sp   & $ $20.4 \\
J123708+621246 &   20 & $>-$0.1 & off  & $ $24.4 \\
J123711+621331 & *~31 & $ -$0.7 & g/b  & $ $23.1 \\
J123716+621512 & *145 & $ -$0.4 & g/b  & $ $20.4 \\
J123721+621129 &  677 & $ -$0.3 & EF   & $>$27.0 \\
J123725+621128 & *613 & $ -$1.2 & g    & $ $24.1 \\
J131157+423910 &  109 & $ -$0.3 & g    & $ $23.5 \\
J131157+423630 &  197 & $ -$0.6 & EF   & $>$25.5 \\
J131203+423331 &  886 & $ -$0.3 & g    & $ $22.2 \\
J131205+423851 &   12 & $ -$0.7 & EF   & $>$25.5 \\
J131207+423945 &   14 & $ -$0.7 & EF   & $>$25.5 \\
J131209+424217 &   93 & $ -$0.4 & g    & $ $23.4 \\
J131211+424053 &  717 & $ -$0.6 & g/b  & $ $20.8 \\
J131213+424129 &   48 & $  $0.0 & Ell  & $ $18.7 \\
J131213+423706 &    9 & $ -$0.8 & g?   & $ $25.5 \\
J131215+423702 &   13 & $ -$0.4 & g/b  & $ $19.7 \\
J131215+423901 & *~27 & $ -$0.5 & Q    & $ $17.8 & YES \\
J131216+423921 &   30 & $ -$0.9 & EF   & $>$25.8 \\
J131217+423912 &   20 & $ -$0.6 & g/b  & $ $20.0 \\
J131217+423930 &   16 & $ -$0.9 & EF   & $>$25.8 \\
J131218+423843 & *~25 & $ -$0.9 & g/b  & $ $20.3 \\
J131219+423608 & *~26 & $ -$0.9 & g    & $ $22.1 \\
J131219+423831 & *~12 & $ -$0.8 & g    & $ $20.4 \\
J131220+423923 &   15 & $ -$0.7 & EF   & $>$25.8 \\
J131220+423704 &   14 & $ -$1.0 & EF   & $>$25.8 \\
J131220+424029 &   34 & $ -$0.3 & EF   & $>$25.5 \\
J131220+423535 &   20 & $ -$0.6 & EF   & $>$25.5 \\
J131221+423923 & *~12 & $ -$0.8 & sp   & $ $18.0 \\
J131221+423722 &   10 & $ -$0.9 & sp   & $ $17.8 \\
J131221+423827 &   19 & $ -$0.8 & g/b  & $ $22.7 \\
J131222+423813 &   10 & $ -$0.5 & Q    & $ $19.6 & YES \\
J131223+423908 & *~19 & $ -$0.8 & g/b  & $ $19.3 \\
J131223+423712 &   28 & $ -$0.8 & g/b  & $ $18.8 \\
J131223+423525 &   66 & $ -$0.1 & Ell  & $ $18.9 \\
J131225+423656 &   13 & $>-$0.1 & g?   & $ $25.0 \\
J131225+424103 &   35 & $ -$0.7 & g?   & $ $25.5 \\
J131225+423941 &  200 & $ -$0.7 & EF   & $>$25.5 \\
J131227+423800 &   26 & $ -$0.9 & g/b  & $ $19.8 \\
J131232+424038 &  166 & $ -$0.8 & g    & $ $24.8 \\
J131239+423911 &   66 & $ -$0.6 & g    & $ $24.5 \\
J171349+501610 & 1138 & $  $    &      & $ $     \\
J171354+501547 &  153 & $  $    &      & $ $     \\
J171411+501602 &   37 & $  $    & g/b  & $ $23,5 \\
J171414+501530 & 6480 & $ -$1.1 & g/b  & $ $22.1 \\
J171415+501535 &   49 & $  $    &      & $ $     \\
J071416+501817 &  495 & $  $    &      & $ $     \\

\enddata

\tablecomments {
* = Source is resolved, $>2.5''$ {\protect \\}
{\bf Galaxy Types} {\protect \\}
{\bf g} - galaxy of unknown type.
{\bf g/b} - interacting galaxies.  Magnitudes given for galaxy associated with the radio source.
{\bf Q} - Quasar.
{\bf g?} - Uncertain identification with a faint object.
{\bf EF} - Empty field with magnitude limit given.
{\bf ell} - Elliptical galaxy.
{\bf sp} - Spiral galaxy.
}	
\end{deluxetable}

\begin{deluxetable}{crcrrc}
\tablecolumns{6}
\tablewidth{0pt}
\tabletypesize{\small}
\tablecaption{The Count of Sources at 8.4 GHz}
\tablehead{
 \colhead {Range of S$_{8.4}$} &
 \colhead {<S$_{8.4}>$} &
 \colhead {Area} &
 \colhead {\# Obs} &
 \colhead {\# Fit} &
 \colhead {Ratio} \\

 \colhead {($\mu$Jy)} &
 \colhead {($\mu$Jy)} &
 \colhead {(arcmin$^2$)} &
 \colhead {} &
 \colhead {} &
 \colhead {Obs/Fit} \\
}
\startdata
  7.5 to  19.0 &  13.7  &  65 & 21 &  26.6 & $0.79 \pm 0.17$ \\
  9.0 to  30.0 &  24.5  & 174 & 20 &  15.6 & $1.28 \pm 0.29$ \\
 30.0 to  65.0 &  44.9  & 317 & 20 &  24.8 & $0.81 \pm 0.18$ \\ 
 65.0 to 600.0 & 195.0  & 558 & 21 &  29.4 & $0.71 \pm 0.16$ \\
$>600.0$       & 1675.0  & 653 &  7 &   3.1 & $2.32 \pm 0.91$ \\

\enddata

\end{deluxetable}


\begin{thebibliography}{}

%\bibitem [Baars et al.\ 1977]{baa77}
%Baars et al.\ 1977

\bibitem [Barger et al.\ 1999]{bar99}
Barger, A.~J., Cowie, L.~L. \& Sanders, D.~B. 1999, \apj, 518 L5

\bibitem [Barger et al.\ 2000]{bar00}
Barger, A.~J., Cowie, L.~L. \& Richards, E.~A. 2000, \aj, 119, 2029

\bibitem [Blain et al.\ 1999]{bla99}
Blain, A.~W., Smail, I., Ivison, R.~J. \& Kneib, J-P. MNRAS, 302, 632

\bibitem [Brant et al.\ 2001]{bra01}
Brandt, W.~N. et al.\ 2001, AJ, 122, 1

\bibitem [Burkey at al.\ 1994]{bur94}
Burkey, J.~M., Keel, W.~C., Windhorst, R.~A. \& Franklin, B.~E. 1994,
\apj, 429, L13

\bibitem [Campos et al.\ 1999]{cam99}
Campos, A., Yahil, A., Windhorst, R. A., Richards, E. A., Pascarelle, S.,
Impey, C., \& Petry, C. 1999, \apj, 511, L001

\bibitem [Condon \& Mitchell 1984]{con84}
Condon, J.~J \& Mitchell, K.~J. 1984, AJ, 89, 610

\bibitem [Condon \& Yin 1990]{con90}
Condon, J.~J \& Yin, Q.~F. 1990, \apj, 357, 97.

\bibitem [Cowie et al.\ 2001]{cow01}
Cowie, L.~L., Barger, A.~J., Bautz, M.~W., et al.\ 2001, \apj, 551, L9

\bibitem [Cram 1998]{cra98}
Cram, L.~E. 1998, \apj, 506, L85

\bibitem [Crawford et al.\ 1970]{cra70}
Crawford, D.~L., Jauncey, D.~L. \& Murdoch, H.~A. 1970, \apj, 162, 405

%\bibitem [Cress et al.\ 1996]{cre96}
%Cress, C. M., Helfand, D. J., Becker, R. H., Gregg, M. D., White, R. L. 1996,
%\apj, 473, 7

\bibitem [Donnelly et al.\ 1987]{don87}
Donnelly, R. H., Partridge, R. B., \& Windhorst, R. A. 1987, \apj, 321, 94

%\bibitem [Driver et al.\ 1995]{dri95}
%Driver, S. P., Windhorst, R. A., Ostrander, E. J., Keel, W. C., Griffiths, R.
%E., \& Ratnatunga, K. U. 1995, \apj, 449, L23

\bibitem [Elbaz et al.\ 1999]{elb99}
Elbaz, D., et al.\ 1999, A\&A, 351, L37

\bibitem [Fomalont et al.\ 1991] {fom91}
Fomalont, E.~B., Windhorst, R.~A., Kristian, J.~A. \& Kellermann,
K.~I. 1991, AJ, 102, 1258

\bibitem [Fomalont et al.\ 1997]{fom97}
Fomalont, E.~B., Kellermann, K.~I., Richards, E.~A., Windhorst, R.~A. \&
Partridge, R.~B. 1997, \apj, 475, L5.

\bibitem [Fomalont et al.\ 2002]{fom02}
Fomalont, E~B., Muxlow, T.~W.~M., et al.\ 2002, in preparation.

\bibitem [Garrett et al.\ 2000]{gar00}
Garrett, M.~A., deBruyn, A~.G., Giroletti, M., Baan, W.~A. \&
Schilizzi, R.~T., 2000, A\&A, 361, L41

\bibitem [Goldschmidt et al.\ 1997]{gol97}
Goldschmidts, P., Oliver, S.~J., Serjeant, S.~B., et al.\ 1997,
MNRAS, 289, 465

\bibitem [Griffiths et al.\ 1994]{gri94}
Griffiths, R. E., et al.\ 1994, \apjl, 435, L19

\bibitem [Haarsma et al.\ 2000]{haa00}
Haarsma, D. B., Partridge, R. B., Windhorst, R. A., \& Richards, E. A. 2000,
\apj, 544, 641

\bibitem [Hammer et al.\ 1995] {ham95}
Hammer, F., Crampton, D., Lilly, S.~J., LeFebre, O. \& Kenet, T.
1995, MNRAS, 276, 1085

\bibitem [Hasinger et al.\ 1998]{has98}
Hasinger, G., Burg, R., Giaconni, R., Schmidt, M., Trumper, J. \&
Zamorani, G. 1998, A\&A, 329, 482

\bibitem [Hopkins et al.\ 1998]{hop98}
Hopkins, A. M., Mobasher, B., Cram, L., \& Rowan-Robinson, M. 1998, MNRAS, 296,
839

\bibitem [Hopkins et al.\ 1999]{hop99}
Hopkins, A., Afonso, J., Cram, L., \& Mobasher, B. 1999, \apj, 519, L59

%\bibitem [Hopkins et al.\ 2000]{hop00}
%Hopkins, A., Windhorst, R. A., Cram, L., \& Ekers, R. 2000, Experimental
%Astronomy, Vol. 10, No. 4, 419

\bibitem [Hughes et al.\ 1998]{hug98}
Hughes, D.~H., Serjeant, S., Dunlop, J., et al.\ 1998, Nature, 394, 241

%\bibitem [Keel et al.\ 1999]{kee99}
%Keel, W. C., Cohen, S. H., Windhorst, R. A., \& Waddington, I. 1999, \aj, 118, 2547

\bibitem [Kellermann \& Richards 1999]{kel99}
Kellermann, K.~I. \& Richards, E.~A., 1999, in Scientific Imperatives at
Centermeter Wavelengths, Ed. M.~P. van Haarlem \& J.~M. van der Huivel,
(Dwingeloo, NFRA), in press

\bibitem [Lilly et al.\ 1996]{lil96}
Lilly, S.~J., LeFevre, O., Hammer, F. \& Crampton, D. 1996, \apj, 460, L1

\bibitem [Madau et al.\ 1996] {mad96}
Madau, P., Ferguson, H.~C., Dickinson, M.~E., Giavalisco, M., Steidel, C.~C.
\& Fruchter, A., 1996, \mnras, 283, 1388

\bibitem [Mobasher et al.\ 1999]{mob99}
Mobasher, B., Cram, L., Georgakakis, A. \& Hopkins, A. 1999, MNRAS,
308, 45.

\bibitem [Mushotzky et al.\ 2000]{mus00}
Mushotzsky, R.~F., Cowie, L.~L., Barger, A.~J. \& Arnaud, K.~A. 2000,
Nature, 404, 459

\bibitem [Muxlow et al.\ 2001]{mux01}
Muxlow, T.~W.~B., Wilkinson, P.~N. et al., in preparation

%\bibitem [Neuschaefer \& Windhorst 1995]{neu95}
%Neuschaefer, L. W., \& Windhorst, R. A. 1995, \apj, 439, 14

%\bibitem [Odewahn et al. \ 1996]{ode96}
%Odewahn et al.\ 1996

\bibitem [Oort \& Windhorst 1985]{oor85}
Oort, M. J. A., \& Windhorst, R. A. 1985, A\&Ap, 145, 405

\bibitem [Partridge et al.\ 1986]{par86} Partridge, R. B., Hilldrup,
K. C., \& Ratner, M. I. 1986, ApJ, 308, 46

\bibitem [Partridge et al.\ 1997] {par97}
Partridge, R.~B., Richards, E.~A., Fomalont, E.~B., Kellermann, K.~I.
\& Windhorst, R.~A. 1997, \apj, 483, 38.

\bibitem [Pascarelle et al.\ 1996a]{pas96a}
Pascarelle, S. M., Windhorst, R. A., Driver, S. P., Ostrander, E. J., \& Keel,
W. C. 1996a, \apj, 456, L021

\bibitem [Pascarelle et al.\ 1996b]{pas96b}
Pascarelle, S. M., Windhorst, R. A., Keel, W. C., \& Odewahn, S. C. 1996b,
Nature, 383, 45

\bibitem [Pascarelle et al.\ 1998]{pas98}
Pascarelle, S. M., Windhorst, R. A., \& Keel, W. C. 1998, AJ, 116, 2659

\bibitem [Pascarelle et al.\ 2001]{pas01}
Pascarelle, S. M., Armus, L., Scoville, N. Z., Windhorst, R. A., \& Cohen, S.
H. 2001, AJ, submitted

\bibitem [Prandoni et al.\ 2001]{pra01}
Prandoni, I., Gregorini, L., Parma, P., de Ruiter, H.~R., Vettolani, G.,
Wieringa, M.~H. \& Ekers, R.~D. 2001, A\&A, 365, 392.

%\bibitem [Richards et al.\ 1997] {ric97}
%Richards et al.\ 1997

\bibitem [Richards et al.\ 1998] {ric98}
Richards, E.~A., Kellermann, K.~I., Fomalont, E.~B., Windhorst, R.~A.
\& Partridge, R.~A. 1998, AJ, 116, 1039

\bibitem [Richards et al.\ 1999] {ric99}
Richards, E.~A., Fomalont, E.~B., Kellermann, K.~I., Windhorst, R.~A.,
Partridge, R.~B., Cowie, L.~L. \& Barger, A.~J. 1999, \apj, 526, L73

\bibitem [Richards 2000]{ric00}
Richards, E.~A. 2000, \apj, 533, 611

\bibitem [Richards et al.\ 2002] {ric02}
Fomalont, E.~B., et al.\ 2002, in preparation

\bibitem[Rowan-Robinson et al.\ 1993]{row93}
Rowan-Robinson, M., Benn, C.~R., Lawrence, A., McMahon, R.~G. \&
Broadhusrt, T.~J. 1993, MNRAS, 263, 123 

\bibitem [Rowan-Robinson et al.\ 1997]{row97}
Rowan-Robinson, M., Mann, R.~G., Oliver, S.~J. et al.\ 1997,
MNRAS, 289, 490

\bibitem [Scott et al.\ 2000]{sco00}
Scott, D., Lagache, G., Borys, C. et al.\ 2000, A\&A, 357, 5

\bibitem [Steidel et al.\ 1999]{ste99}
Steidel, C.~C., Adelberger, K.~L., Giavalisco, M., Dickinson, M. \&
Pettini, M. \apj, 519, 1

\bibitem [Waddington et al.\ 2001]{wad01}
Waddington, I., Windhorst, R.~A., Dunlop, J.~S., Peacock, J~.A. \&
Koos, D.~C. 2001, MNRAS, in press

%\bibitem [Weistrop et al.\ 1987] {wei87}
%Weistrop et al.\ 1987

\bibitem [Williams et al.\ 2000]{wil00}
Williams, R.~E., et al.\, 2000, AJ, 120, 2735

\bibitem [Windhorst et al.\ 1984]{win84}
Windhorst, R.~ A., van Heerde, G., \& Katgert, P. 1984, A\&A Sup. 58, 1

\bibitem [Windhorst et al.\ 1985]{win85}
Windhorst, R. A., Miley, G. K., Owen, F. N., Kron, R. G., \& Koo, D. C. 1985,
\apj, 289, 494

\bibitem [Windhorst et al.\ 1993]{win93}
Windhorst, R.~A., Fomalont, E.~B., Kellermann, K.~I., Partridge, R.~B.
\& Lowenthal J.~D. 1993, ApJ, 405, 498

\bibitem [Windhorst et al.\ 1994]{win94}
Windhorst, R. A., et al.\ 1994, AJ, 107, 930

\bibitem [Windhorst et al.\ 1995]{win95}
Windhorst, R.~A., Fomalont, E.~B., Kellermann, K.~I., Partidge, R.~B.,
Richards, E.~A., Franklin, B.~E., Pascarelle, S.~M. \& Griffiths, R.~E.
1995, Nature, 375, 471

\bibitem [Windhorst et al.\ 1998]{win98}
Windhorst, R. A., Keel, W. C., \& Pascarelle, S. M. 1998, \apj, 494, L27

\end{thebibliography}
\end{document}